\documentclass[submission,copyright,creativecommons]{eptcs}
\providecommand{\event}{NCMA 2022} 
\usepackage{breakurl}             
\usepackage{underscore}           
\usepackage{graphicx}
\usepackage{wrapfig}
\usepackage{amsmath}
\usepackage{listings}

\usepackage{xcolor}
\usepackage[linguistics]{forest}
\usetikzlibrary{decorations.pathreplacing}

\newcommand{\where}{\:|\ }

\title{Practical Aspects of Membership Problem of Watson-Crick Context-free Grammars}
\author{Jan Hammer
\institute{Faculty of Information Technology\\
Brno University of Technology\\
Czech Republic}
\email{xhamme00@stud.fit.vutbr.cz}
\and
Zbyněk Křivka
\institute{Faculty of Information Technology\\
Brno University of Technology\\
Czech Republic}
\email{krivka@fit.vutbr.cz}
}
\def\titlerunning{Membership Problem of WK Context-free Grammars}
\def\authorrunning{J. Hammer, Z. Křivka}

\newcommand{\wkdomain}[2]{\big[\genfrac{}{}{0pt}{1}{#1}{#2}\big]}
\newcommand{\wkpair}[2]{\big(\genfrac{}{}{0pt}{1}{#1}{#2}\big)}

\begin{document} 
\maketitle

\begin{abstract}
This paper focuses on Watson-Crick languages inspired by DNA computing, their models, and algorithms for deciding the language membership. It analyzes a recently introduced algorithm called WK-CYK and introduces a state space search algorithm that is based on regular Breadth-first search but uses a number of optimizations and heuristics to be efficient in practical use and able to analyze longer inputs. The key parts are the heuristics for pruning the state space (detecting dead ends) and heuristics for choosing the most promising branches to continue the search.

These two algorithms have been tested with 20 different Watson-Crick grammars (40 including their Chomsky normal form versions). While WK-CYK is able to decide the language membership in a reasonable time for inputs of the length of roughly 30--50 symbols and its performance is very consistent for all kinds of grammars and inputs, the state space search is usually (89--98 \% of cases) more efficient and able to do the computation for inputs with lengths of hundreds or even thousands of symbols. Thus, the state space search has the potential to be a good tool for practical Watson-Crick membership testing and is a good basis for improvement the efficiency of the algorithm in the future.
\end{abstract}

\section{Introduction}

The ability to read DNA, to understand it or even to modify it, is certainly one of the ways that many people think will define the future. But in order to work with DNA, there needs to be a mathematical model that can actually do calculations with such structures and that is prepared to be run on computers. Moreover, working with this model must be efficient enough because the length of string representation of genetic code is usually huge.

This work follows the work of M. Zulkufli et al. \cite{WK-GRAMMARS-1}, \cite{WK-GRAMMARS-2}, \cite{WK-CYK} who have studied models for working with Watson-Crick languages and introduced the WK-CYK algorithm, a modification of the CYK algorithm, which works with Watson-Crick context-free grammars and is able to decide the membership problem for these languages. The stated complexity of this algorithm is $\mathcal{O}(n^6)$ with respect to the input length. However, with this complexity, the algorithm still does not seem to be useful for practical DNA computations, considering how long the DNA code is.

Therefore, this paper introduces the state space search algorithm. While its theoretical complexity is not as good as in the case of WK-CYK, it takes a more practical approach. In practice, thanks to various heuristics, it is very often able to decide the membership in languages defined by Watson-Crick context-free grammars of inputs far longer than what WK-CYK can handle on today's computers.

\section{Definitions}

Several models working with double-stranded sequences inspired by DNA (\cite{DNA-computing}) have been proposed. The description here is focused on the Watson-Crick (WK) context-free grammars, the model that both the algorithms discussed in this paper work with.

\subsection{Watson-Crick grammars}

The first kind of Watson-Crick grammars (WK grammars) introduced were the WK regular grammars \cite{REG-GRAMMAR}. The key features are shared with Watson-Crick automata (introduced in \cite{WK-FIN-AUT}). Specifically, it is the complementarity relation $\rho$ and the double-stranded strings that the grammar produces.
The WK regular grammars have been used as a basis for Watson-Crick linear grammars and Watson-Crick context-free grammars introduced in \cite{WK-GRAMMARS-1}.

A \emph{Watson-Crick context-free grammar} is $G = (N, T, \rho, P, S)$ where $N$ is a finite set of non-termi\-nals, $T$ is a finite set of terminals and $N \cap T = \emptyset$, $S \in N$ is a starting non-terminal, $\rho \subseteq T \times T$ is a symmetric complementarity relation, and $P$ is a finite set of rules that have the form $A \rightarrow \alpha$ where $A \in N$, $ \alpha \in (N \cup \wkpair{T^*}{T^*})^*$. $\wkpair{w_1}{w_2}$ denotes simply a pair $(w_1, w_2)$. Set $V = N \cup T$.

A \emph{Watson-Crick domain} (WK domain) denotes all valid double strands associated with given $T$ and $\rho$. Formally:
$$\wkdomain{T}{T}_{\rho}^* = \Big\{\wkpair{a}{b} \where a, b \in T\textnormal{, }(a, b) \in \rho \Big\}^*$$
This implies that both strands in a WK domain must have the same length.

The derivation by the grammar $G$ starts with the starting symbol $S$. For $x, y \in (N \cup \wkpair{T^*}{T^*})^*$ $x$ directly derives $y$, denoted by $x \Rightarrow y$, if and only if $x = \beta A \gamma$ and $y = \beta \alpha \gamma$
where $A \in N$, $\alpha, \beta, \gamma \in (N \cup \wkpair{T^*}{T^*})^*$, and $A \to \alpha \in P$.
The language generated by $G$ is defined as 
$L(G) = \big\{ w_1| S \Rightarrow^* \wkpair{w_1}{w_2}\textnormal{ and } \wkpair{w_1}{w_2} \in \wkdomain{T}{T}_{\rho}^* \big\}$
where $\Rightarrow^*$ is a reflexive and transitive closure of $\Rightarrow$.

The symbol $\lambda$ denotes the empty string. A $\lambda$-rule is a rule of the form $A \rightarrow \wkpair{\lambda}{\lambda}$, i.e., a rule that simply removes the left-hand side non-terminal (while producing the empty string). A \emph{double-stranded string} (DS-string) is a pair from $\wkpair{T^*}{T^*}$. A \emph{letter}, $X$,  is a non-terminal or a DS-string, so $X \in N \cup \wkpair{T^*}{T^*}$ and a \emph{Watson-Crick word} (WK-word) is a string of letters from $\big(N \cup \wkpair{T^*}{T^*}\big)^*$.

\subsection{Expressive power of Watson-Crick models}
\begin{wrapfigure}{R}{0.5\textwidth}
   \includegraphics[height=5.5cm]{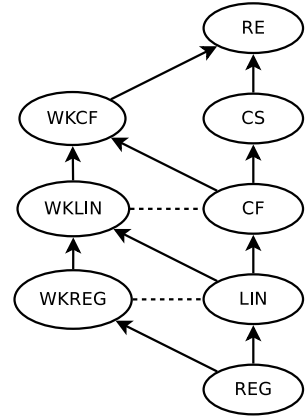}
   \centering
   \caption{Comparison of WK language families in the context of the Chomsky hierarchy. The full arrows denote proper inclusion, dotted arrows denote inclusion and dotted lines denote incomparability.}
   \label{fig:expr-power}
\end{wrapfigure}

The comparison of the expressive power of WK language families and the Chomsky hierarchy studied in \cite{WK-GRAMMARS-1} and \cite{WK-GRAMMARS-2} is recalled in Figure \ref{fig:expr-power},
 where the Chomsky hierarchy is visualized on the right (\textbf{REG}, \textbf{LIN}, \textbf{CF}, \textbf{CS} and \textbf{RE} denote regular, linear, context-free, context-sensitive and recursively enumerable languages, respectively) while the Watson-Crick languages are on the left. \textbf{WKREG}, \textbf{WKLIN}, and \textbf{WKCF} denote the families of languages defined by Watson-Crick regular grammars, WK linear grammars, and WK context-free grammars. Note that, by \cite{REG-GRAMMAR}, non-deterministic WK automata define \textbf{WKREG} as well. It is still an open problem whether WK pushdown automata \cite{WK-PUSHDOWN} define \textbf{WKCF}.

It has been shown in \cite{COMPL-REL} that the type of complementary relation which is used does not increase the expressive power of WK automata and grammars. Moreover, \cite{SURVEY} provides an algorithm to transform any WK automaton into an equivalent WK automaton with the relation being identity. Therefore, many models and algorithms limit themselves to work only with identity complementarity relation.

\pagebreak[4]
\section{Remarks on WK-CYK algorithm}
So far, the WK-CYK algorithm introduced in \cite{WK-CYK} (from where it is taken and attached to Appendix A) has been practically the only algorithm explicitly designed to decide a membership in Watson-Crick languages defined by a WK context-free grammars. It is an enhancement of the CYK algorithm \cite{CYK} modified for WK context-free languages.

This algorithm works with grammars in Watson-Crick Chomsky normal form (WK-CNF), a modification of the Chomsky normal form for context-free grammars. All rules in WK-CNF must be in one of the following forms:

\begin{itemize}
  \item{$A \rightarrow \wkpair{a}{\lambda}$}
  \item{$A \rightarrow \wkpair{\lambda}{a}$}
  \item{$A \rightarrow B C$}
  \item{$S \rightarrow \wkpair{\lambda}{\lambda}$ (this rule is used only to include the empty string in the language)}
\end{itemize}

where $A$, $B$ and $C$ are non-terminals, $S$ is the starting non-terminal and $a$ is a terminal of the grammar. It is possible to transform any WK context-free grammar into the WK-CNF grammar as described in detail in \cite{WK-CYK}. The WK-CYK algorithm is further analyzed in \cite{DIPL} with the following two remarks worth noting.

1. The loop on Line 9 of \textit{SetsConstruction} procedure (see Appendix A) iterates $\beta$ from 0 to $n$. In this context, $\beta$ represents the length of the lower strand while $\alpha$ represents the length of the upper strand of a particular DS-string sub-sequence and $\alpha = y - \beta$ where $y$ is the length of the DS-string sub-sequence. Part of the loop actually calculates with nonsensical values. When calculating with a sub-sequence that is shorter then one strand (i.e. $y < n$), it includes the case when $\beta > y$ so $\alpha < 0$. For instance, the algorithm splits the sub-sequence of length 2 into two parts of lengths 3 and -1.

When calculating with a subsequence that is longer than the input, it includes the case where $\beta$ is too short, so $\alpha$ is then longer than a strand length. For instance, if the input length is 4 and the sub-sequence length is 8, it splits the sub-sequence to lengths 7 and 1, which is not possible with the input length of 4.

This does not affect the correctness of the computation because the nonsensical values find no result. However, a more precise and efficient solution is to iterate $\beta$ over the interval $[max(y-n, 0)$, $min(n, y)]$ instead of the interval $[0$, $n]$. This modification is included in our implementation.

\medskip

2. The time complexity of WK-CYK is $\mathcal{O}(n^6)$. As described in \cite{WK-CYK} (Section 6), the WK-CYK main procedure has the time complexity of $\mathcal{O}(n^4)$ and the nested \textit{ComputeSet} procedure has the time complexity of $\mathcal{O}(n^2)$. This is true with respect to the input length. Possibly, a more precise description of the complexity would be $\mathcal{O}(n^6 \times R)$, where $n$ is the input length and $R$ is the number of rules in the grammar. The description of \textit{ComputeSet} procedure  uses the operation of set union ($\cup$), as if it has constant time complexity, which, in reality, it does not---it requires iterating over the rules of the grammar.

\section{Testing membership by searching the state space}

This section introduces the main algorithm of this paper for testing membership in WK context-free languages. In the following sections, it is referred to as the \emph{state space search} or the \emph{tree search}. Its core is a standard Breadth-first search algorithm (BFS) with various optimizations added on top.

Standard BFS starts with a root node. In the case of grammar, that is the starting non-terminal symbol of the grammar. Then the tree is built by applying all possible rules to all possible non-terminals. Each rule application generates a new node. The node contains a WK-word, which consists of some non-terminals, some terminals in the upper strand and some terminals in the lower strand.

The BFS algorithm always finds a solution if there is one. It finds the optimal solution, which, in this case, means the shortest sequence of rules that generate the input string from the starting non-terminal. However, whether the solution is optimal or not is irrelevant to the membership problem. If there is no solution, the algorithm will probably never stop, as the state tree is usually infinite. Also, such a tree would grow very rapidly and the solution would usually not be found in a reasonable time frame. Therefore, some optimizations need to be used. This work introduces two key kinds of optimizations. Firstly, identifying dead ends in the search tree and removing them from the computation---this is referred to as \emph{pruning}. Secondly, choosing such nodes for the subsequent computation, which seem to be the most promising in leading to the solution. This is referred to as \emph{node precedence}.

\subsection{Key characteristics of the state space search}
Besides pruning and node precedence heuristics, the algorithm keeps a set of states, which have been generated (added to the tree), in order to avoid analyzing the same WK-word repeatedly or even getting stuck in a loop. Also, it considers left derivation only. This means that a node that contains several non-terminals can generate new nodes only by applying rules to the first non-terminal in the WK-word.

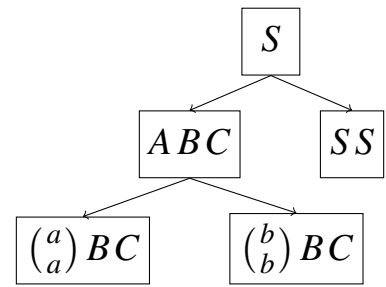
\begin{wrapfigure}{r}{0.45\textwidth}
   \centering
   \begin{forest}
    for tree={rectangle, black, draw, minimum size = 2em, font=\Large, edge={->}, s sep = 30pt}
    [$S$,name=root, edge={->}
        [$A\:B\:C$, name=bordertop, edge={->}
            [$\wkpair{a}{a}\:B\:C$, name=borderleft]
            [$\wkpair{b}{b}\:B\:C$,name=borderbottom]
        ]
        [$S\:S$,name=borderright]
    ]
    \end{forest}
   \caption{Example of a search tree}
   \label{fig:search-tree}
\end{wrapfigure}
Figure \ref{fig:search-tree} shows an example of a tree search progress. The rules of its grammar in this example are $S \rightarrow S S \:|\: A B C$, $A \rightarrow \wkpair{a}{a} \:|\: \wkpair{b}{b}$ and some rules $B \rightarrow ...$, $C \rightarrow ...$ that are not important. $S$ is the starting non-terminal, therefore, $S$ is the first node and there are two possible rules that can be applied to $S$, so this node has two successors. The node precedence heuristic will choose one of the successors to be analyzed next---perhaps the left one with the WK-word $A B C$. This node, too, only has two successors, which are made by the two rules that can be applied to the first non-terminal, $A$. Even though there are some rules for $B$ and $C$, these rules are not used to produce successors, yet. The nodes created by rules applied on $B$ would be successors of the WK-words $\wkpair{a}{a} B C$ and $\wkpair{b}{b} B C$, which have the symbol $B$ as the first non-terminal from the left.

In a WK-word of a WK grammar, the terminals are clustered together into DS-strings. If two DS-strings appear next to each other in a WK-word, they are merged. These DS-strings, as well as non-terminals, are referred to as letters because together they constitute words. For instance, a three-letter WK-word: $\wkpair{abc}{\lambda} A \wkpair{b}{b}$ after application of rule $A \rightarrow \wkpair{\lambda}{a}$, will result in a WK-word with just one letter: $\wkpair{abcb}{ab}$.

The WK-word in a node that is the solution needs to meet the following criteria:
\begin{enumerate}
  \item{It contains no non-terminals. Since neighboring DS-strings are always merged, this implies that there is only one letter---a DS string.}

  \item{The upper and lower strands of the DS-string are of the same length.}

  \item{Each pair of symbols from the upper and lower strands with the same index must be related by the complementarity relation.}

  \item{The upper strand must be equal to the input string.}
\end{enumerate}

If the criteria are met, the algorithm has found the right node and that means the input string is an element of the language defined by the grammar. It has been accepted by the state space search algorithm. If the whole state space has been searched (in case it is not infinite), there is no solution and the input has been rejected by the state space search algorithm.

\subsection{Identifying a dead end in the state tree}
A blind BFS would stop searching a branch only when all non-terminals have been used to generate all possible terminal WK-words (WK-words with terminals only). But sometimes it is possible to tell in advance that a specific WK-word cannot lead to the desired solution. If that is the case, the node can simply be removed and the whole branch which it would generate is skipped. The next section describes various ways (heuristics) of recognizing the dead branches. These are referred to as pruning heuristics, there are five of them and each one has an abbreviation which is used further on.

\begin{enumerate}
  \item{Detecting that one of the strands is already too long (SL)}
  \item{Detecting that the overall WK-word is already too long (TL)}
  \item{Matching the starting terminals in the upper strand to the input (WS)}
  \item{Checking the complementarity relation (RL)}
  \item{Comparing the input to a regular expression generated from the WK-word (RE)}
\end{enumerate}

\subsubsection{One of the strands is too long (SL)}
A terminal symbol which appears either in the upper or lower strand can never disappear further in the branch. That means that the count of all symbols in the upper and in the lower strand must not be greater than the length of the input string. Otherwise, the solution can never be reached from that branch.

\subsubsection{The WK-word including non-terminals is too long (TL)}
Non-terminals present a more complex problem when dealing with the length of the WK-word. First of all, the algorithm calculates in advance how many terminals each non-terminal produces at minimum. For instance, if the grammar contains rules: $A \rightarrow A A \:|\: \wkpair{ab}{cd} \:|\: B B$ and $B \rightarrow \wkpair{a}{\lambda}$, the non-terminal $B$ always produces one terminal, which means one terminal at minimum. The non-terminal $A$ can produce various numbers of terminals, but two at minimum---thanks to the rule $A \rightarrow B B$ and the fact that $B$ has the minimum of one. This value is then considered to be the length of the given non-terminal.
This length can be applied both to the upper or to the lower strand because, in general, it is not known which strand will absorb the symbols generated from the non-terminal.
This then leads to the following constraint on the WK-word:

$$|upper| + |lower| + |nts| \leq 2 \times |input|$$

where $|upper|$ and $|lower|$ are the counts of terminals in the upper and lower strands, $|nts|$ is the length of all non-terminals in the WK-word and $|input|$ is the length of the input string. If this constraint is broken, the WK-word cannot lead to the solution and the branch can be pruned.

If the grammar contains no $\lambda$-rule (rule of the form $N \rightarrow \wkpair{\lambda}{\lambda}$), This constraint guarantees that the algorithm will finish. Once all the WK-words within the given length limit have been generated and a solution has not been found, the search will end.

If the grammar does contain $\lambda$-rules, the previous constraint can still be applied---the non-terminals that can be erased are assigned the length of zero. In this case, it is not possible to guarantee that the search will end, because the non-terminals of length zero can be combined infinitely many times. However, it is possible to utilize the algorithm for removing $\lambda$-rules (described in \cite{WK-CYK}).

\subsubsection{The beginning of the WK-word does not match the input (WS)}
If a WK-word in a node begins with some terminal symbols in the upper strand, these symbols will always stay at the beginning further in the given branch. Unlike the other terminals, these starting terminals already have fixed indexes. If these symbols do not match the prefix of the input string of the same length, the input string can never be generated from this branch.
If, on the other hand, the WK-word starts with a non-terminal, there is nothing to be said about what can be at the beginning of the WK-word further in the branch.

It is possible to check the end of the WK-word in the same manner, but the generation is performed from left to right and so there is little benefit in checking the end.

\subsubsection{Checking the complementarity relation (RL)}
As previously described, the symbols in the upper and lower strands with the same indexes must be related by the complementarity relation. Unfortunately, this can be checked only at the beginning of the WK-word (Technically, it can be checked at the end as well, while indexes of these symbols are not yet known, the last terminal symbol will always stay the last. But just like in the case of the previous heuristic, there is little benefit in checking the end when the generation is done from the left side.). Indexes of the symbols in the middle part (anywhere after the first non-terminal) are not known. Thus this check can be understood as an extension of the previous one---if the WK-word begins with some terminal symbols and there are some symbols in both the upper and lower strands, these symbols can be tested whether they adhere to the complementarity relation. But only to the length of the shorter of the two strands in this letter.

\subsubsection{The input matches a regular expression generated from the WK-word (RE)}
It is possible to generate a regular expression that represents the current WK-word. Each non-terminal serves as a wild card (\verb/.*/). Each terminal in the upper strand stands for itself. The lower strand is ignored. This expression must be matchable to the input string. Otherwise, it is not possible to generate it from the current branch. For instance, if the WK-word is
$$\wkpair{abc}{f}N_1\wkpair{d}{gh}N_2\wkpair{e}{i}N_3$$

then the resulting regular expression will be: $\verb/^abc.*d.*e/$. The symbols $abc$ must be at the beginning (therefore, the \verb/^/ denoting the beginning of the expression is placed at the start); then it is not known what will be generated by the non-terminal $N_1$---therefore, the wildcard is there next; then there will have to be a symbol $d$; another wildcard for non-terminal $N_2$; symbol $e$; and then anything. The regular expression might end with a wildcard generated from the last non-terminal $N_3$, but that is not necessary. Wildcard before and after the expression is implicit. A starting non-terminal can be represented by omitting the symbol \verb/^/ which denotes the beginning of the string. An ending non-terminal can be represented by omitting the symbol \verb/$/ which denotes the end of the string.

\medskip

The order in which the pruning heuristics are applied matters. It is good first to apply the heuristics that are more likely to succeed and that require less computational power. If they are successful, the more complex heuristic can be skipped.

It is possible to come up with some more checks that could identify a dead end in the search tree. The disadvantage of any check is the computing power that has to be used for checking any node that is generated and analyzed. If some checks are unlikely to significantly prune the tree and/or are complicated to compute, it is not clear if they will improve the actual performance of the algorithm.

\subsection{Heuristics for node precedence} \label{heur-node-pref}

The aim of the node precedence heuristics is to choose a path in the search tree, which is likely to lead to the solution, the more promising nodes are taken before the others and their successors are generated sooner. The individual heuristic functions attempt to answer the question---which node is more promising than the rest? It assigns each node a number---an evaluation of the node. The lower the node evaluation, the higher priority the node has.

Such heuristics can only be effective if the answer to the search is positive---if there actually is a solution. Unfortunately, if it is negative, it does not help that the algorithm eliminates the more promising branches of the tree first. Eventually, it will have to search through all possible states anyway, in order to make sure that there is no solution.

The following node precedence heuristics have been implemented. Each heuristic also has an abbreviation that will refer to it further on.

\begin{itemize}
  \item{No heuristic (NONE) -- the evaluation of the WK-word is always 0. This is used for comparison to the other heuristics.}
  \item{Aversion to non-terminals (NTA) -- the evaluation is equal to the count of non-terminals in the WK-word}
  \item{Weighted aversion to non-terminals (WNTA) -- each non-terminal has a pre-calculated weight, which is the minimum amount of rules that must be used in order to generate a terminal WK-word from it. The evaluation is equal to the sum of the weights of all non-terminals in the WK-word.}
  \item{The terminal matching -- there are three variants that differ slightly (TM1, TM2, TM3). Each of them increases the priority (i.e. decreases the evaluation) for each upper strand non-terminal (going from left to right) which matches the input string symbol on the same index.
  \begin{itemize}
    \item{TM1 examines terminals going from the left while ignoring non-terminals, decreases evaluation (i.e. increases priority) for each match and finishes when it discovers the first difference.}
    \item{TM2 is similar to TM1, but when it discovers a difference, it does not finish but increases the evaluation and moves on}
    \item{TM3 evaluates the first item in the WK-word only. If it is a non-terminal, it returns zero.}
  \end{itemize}
  }
  \item{Combinations of NTA/WNTA and TM1/TM2/TM3 -- There are six combinations because it does not make sense to combine NTA and WNTA or TM1-3 together: NTA+TM1, NTA+TM2, NTA+ TM3, WNTA+TM1, WNTA+TM2, WNTA+TM3.}
\end{itemize}

In summary, there are 12 node precedence heuristics considered in total (including the first, empty heuristic). Unlike in the case of pruning, where all methods can be applied at the same time, there can be only one node precedence heuristic active at one time. Therefore Section \ref{section:testing} contains the tests and comparison of the effectiveness of these heuristics.

\subsection{Theoretical complexity of the state space search}
The state space search algorithm uses Breadth-first search (BFS) as its basis. Both the time and space complexity of BFS are $\mathcal{O}(b^d)$, where $b$ is the maximum number of successors of a node (branching factor) and $d$ is the depth of the tree. The branching factor is then equal to the maximum number of rules of the given grammar that have the same non-terminal on the left-hand side. This is because always only the first non-terminal in the WK-word is used to generate successors in the tree. The depth of the tree is going to be different for different grammars and even for different inputs.

In general, the theoretical complexity of the state space search algorithm is not impressive, it is much worse then WK-CYK's $\mathcal{O}(n^6)$ or $\mathcal{O}(n^6 \times R)$. However, this is because it has been designed with a rather practical approach, it relies heavily on the heuristics and optimizations, so its performance is usually much better.

\section{Remarks on the implementation of the state space search}
Regarding the current implementation of the state space search, the authors have used Python programming language to keep the implementation simple.\footnote{Our implementation is available at \url{https://github.com/xhamme00/NCMA22_WK_models}}  Next, we emphasize several implementation aspects.

\paragraph{Pruning and node precedence should be flexible}
An important feature of the state space search is the possibility to choose and switch node precedence functions and to turn on and off the pruning functions. Therefore, a suitable structure should be used. For the node precedence, a list of node precedence methods and an index of the currently active one works well. The function that calculates the node precedence simply calls the right method from the list on the current index. Switching the node precedence then simply means changing the index. The pruning methods can have a dictionary with pruning functions as keys and Boolean values that specify whether the given pruning is active. Then, the pruning function iterates over this dictionary and calls all active pruning sub-functions.

\paragraph{Rules should be quickly accessible} It is ineffective to store rules in a list. A usual use case is finding all rules that have a specific non-terminal on the left-hand side. Therefore, a dictionary with a non-terminal as a key and a list of rules as a value is suitable. Similarly, if the grammar should have a more complex complementarity relation, it may be suitable to create a dictionary with a terminal symbol as a key and all related terminal symbols (in a list or a string) as a value.

\paragraph{Priority queue}
A suitable way to implement the queue of the open nodes of the BFS tree is a priority queue (python \textit{PriorityQueue} from the standard \textit{queue} module). Every time a node is created, its node precedence evaluation is calculated by the active function (lower result means higher priority). 

\paragraph{Set of generated nodes} It is important to keep track of all the nodes that have already been generated. It is not necessary to remember the nodes themselves, their hash code is enough (unless the search should reveal the path to the solution, if it is found). The hashes are not only smaller but also much faster to test for equality. Every time a node is to be added to the queue, it is first checked whether its hash is in the set of all hash codes of the generated nodes. In the negative case, the node is added to the priority queue and its hash code into this set.

\paragraph{The length of rules and non-terminals}
One of the most important pruning heuristics (TL) works with lengths of non-terminals. This is the minimum amount of terminal symbols that can be generated from a non-terminal. This should be calculated in advance for the whole grammar. Also, the length of rules should be calculated in advance. The length of a rule indicates how the length of a WK-word changes after the rule is applied. The length of a rule is equal to the length of its right-hand side (number of all terminal symbols plus length of all non-terminals) minus the length of the left-hand side non-terminal. If this is pre-calculated, the length of new WK-words can be calculated as the length of its predecessor plus the length of the rule being applied.

\paragraph{Pre-calculating non-terminal distances}
The WNTA node precedence works with distances of non-terminals. This is a minimal number of rules that need to be used before the non-terminal is transformed to a terminal string. This also can be pre-calculated in advance.

\section{Testing the state space search and the WK-CYK algorithm} \label{section:testing}
The state space search and the WK-CYK algorithms have been tested using 20 WK-context-free grammars (see appendix). The grammars are used both in their basic form and after transformation to the WK-CNF which effectively leads to 40 grammars. The testing has been done in the following stages:

\begin{enumerate}
  \item{comparison of the node precedence heuristics and analysis of their efficiency,}
  \item{comparison of the pruning heuristics and analysis of their efficiency,}
  \item{analysis of the time complexity based on the length of the input string, and}
  \item{testing the WK-CYK algorithm with various grammars and comparison to the state space search.}
\end{enumerate}

Throughout all these tests, it is important to keep in mind the following caveats. The generation of input strings contains random elements. They are generated in a way that can specify whether they should be accepted by the grammar or not and that specify the input length, but other than that, the inputs are generally random. In case of some grammars, the specific input generated may have some impact on the complexity of the search. And even with the same input, the search time has some volatility. This is because it is not specified in what order the successors of a node should be generated (i.e. in what order the rules of the grammar should be applied). In most cases, these differences are negligible.

\subsection{Comparison of the node precedence heuristics efficiency} \label{section:node-prec-test}
In Section \ref{heur-node-pref}, 12 node precedence heuristics have been described and only one of them can be active at a time. In order to test their effectiveness, one test has been run for all 40 grammars. It is not useful to test the node precedence heuristics with inputs that are not within the given language as in such cases, the whole space state needs to be searched and node precedence cannot help in any way. The input strings have been chosen to have suitable lengths, so that the computation is finished (at least with some heuristics) in a reasonable time, specifically within the time limit of ten seconds, but the search also should last some measurable amount of time. Each of the 40 tests consists of 12 runs, each with a different node precedence heuristic active.

There are three metrics to observe:
\begin{itemize}
  \item{How many times did the search time exceed the 10-second limit?}

  \item{What is the total time in which all 40 tests were completed for each of the heuristics. There should be a kind of penalty involved if the test times out because the time needed for the computation is, in that case, certainly greater than the time it actually ran before it was stopped by the time limit. Therefore, for the sake of the comparison, the time of the search is in this case doubled.}

  \item{The total time normalized for each test -- all the times are multiplied by a number $n = 1 / t_{min}$ where $t_{min}$ is the time of the fastest heuristic for the given test. This is probably the most telling metric as each test has roughly the same impact on the final number.}
\end{itemize}

\begin{figure}[h!]
  \includegraphics[width=\textwidth]{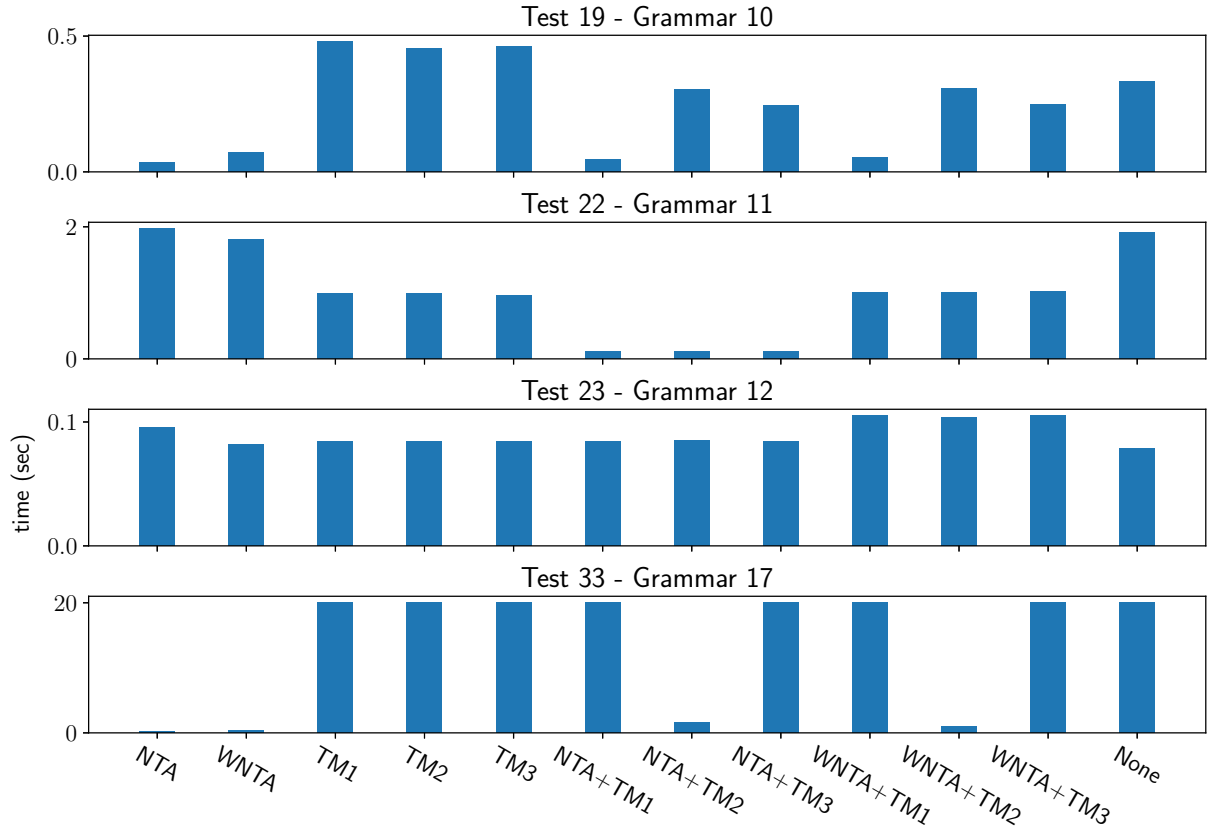}
  \caption{Selected comparisons of node precedence heuristics}
  \label{fig:selected-tests}
\end{figure}

It is interesting to notice how different heuristics are better in different test cases. This is illustrated by selected test cases that are in Figure \ref{fig:selected-tests}. There are some cases where the best heuristic is the empty one which assigns zero to each node, like in the case of test 23. This is because this heuristic is the simplest one to compute and if no heuristic is effective in a particular test case, this one wins. But since it does not win by a large margin, these cases do not have a decisive impact on the overall result. There was only one timeout of heuristic TM2 and the empty heuristic. The significant result differences indicate that the state space search can be customized to fit a specific grammar and thus further improve its performance.

In some cases, certain heuristics do not work so well, but their combination does. This can be seen in test 22 -- NTA and WNTA have poor result, comparable to no heuristic. TM1, TM2, and TM3 have a somewhat better result, but by far, the best result is achieved by a combination of NTA with any version of TM.

\begin{figure}[h!]
  \includegraphics[width=\textwidth]{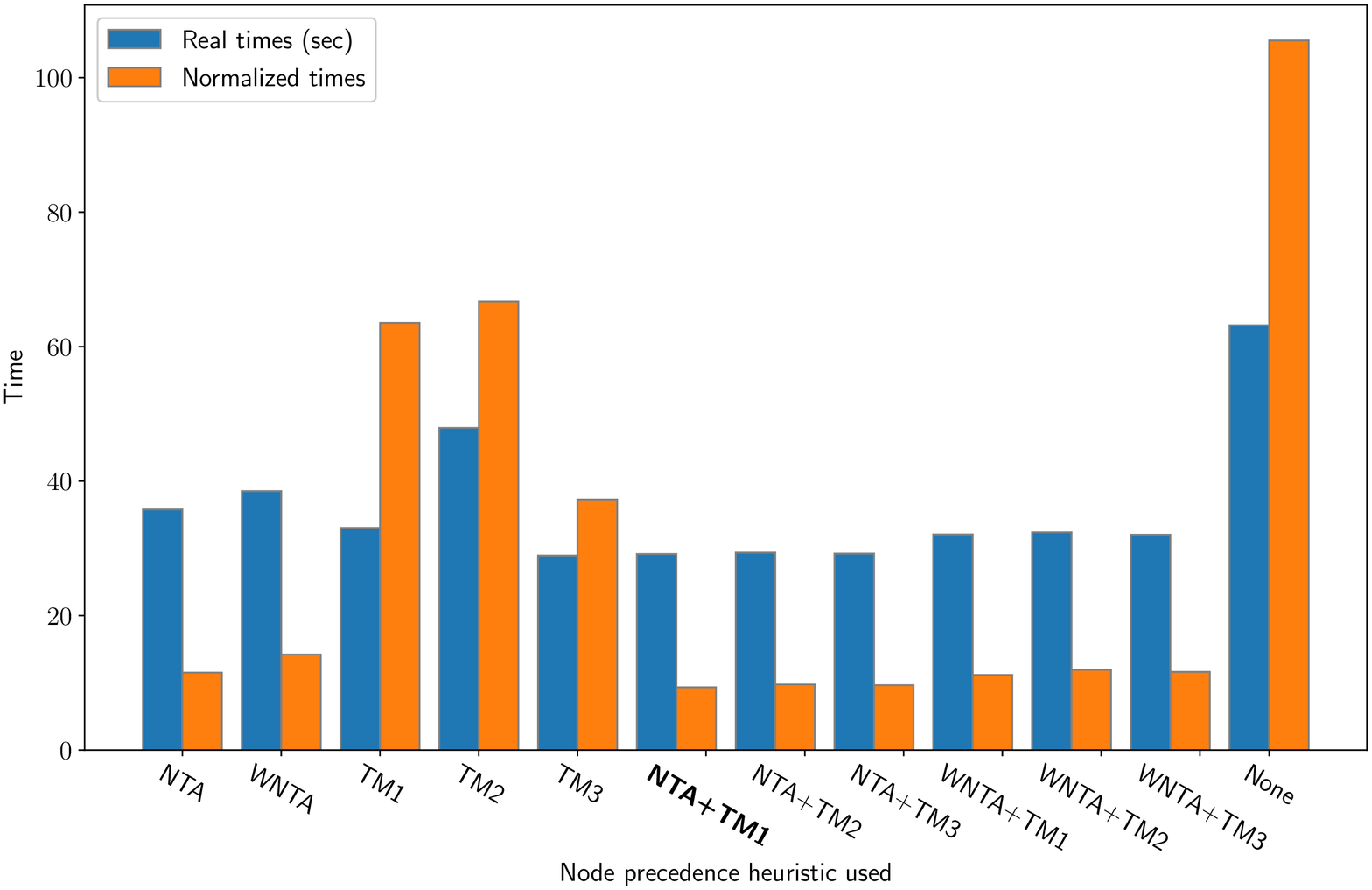}
  \caption{Comparison of the node precedence heuristic functions}
  \label{fig:node-heuristics-comp}
\end{figure}

Figure \ref{fig:node-heuristics-comp} shows the total result for all 40 tests. The left bar of each heuristic shows the total time for all 40 tests and the right bar shows the normalized time. It turns out that the best results are achieved by the combination NTA+TM1. The best time overall was achieved by TM3 by a very narrow margin. The normalized results of all NTA+TM1, NTA+TM2 and NTA+TM3 heuristics are very close, but the narrow winner is NTA+TM1. Even if, in some cases, there are some faster heuristics, it is usually close. Interestingly, even though TM2 turns out to have the worst results of them all, often worse than no heuristic, with the combination of NTA, the results are among the best. Anyway, for all of the following tests, the winning node precedence heuristic NTA+TM1 will be used.

\subsection{Comparison of the pruning heuristics efficiency}
Pruning has the advantage of being useful whether the input string is going to be accepted or rejected by the tree search. Also, all of the pruning can be active at the same time. Each node can be tested by all available checks to see whether it can be pruned or not.

The testing is performed over 80 tests---each of the 40 grammars is used for a positive test (where the input will be accepted) and a negative test (the input will be rejected).
Each test contains seven runs of the tree search algorithm---one where all pruning heuristics are active, one where all are inactive, and one for each heuristic where all are active except the given one.

Similarly to the node precedence heuristics comparison, the metrics that are important are the total time needed to compute the 80 tests and the number of timeouts for each of the seven cases. The main goal here is not to compare the heuristics to each other and find which one is the best---as they can be active at the same time, it does not matter that much. Rather, the goal is to decide whether each of the pruning heuristics improves the performance or if it is better to turn some off. Therefore the comparison between the case where all heuristics are active and the case where a specific heuristic is inactive and the rest are active is important. If the latter case is faster, the given heuristic cost (in terms of computing power) is greater than its contribution.

\begin{figure}[h!]
  \centering
  \includegraphics[width=\textwidth]{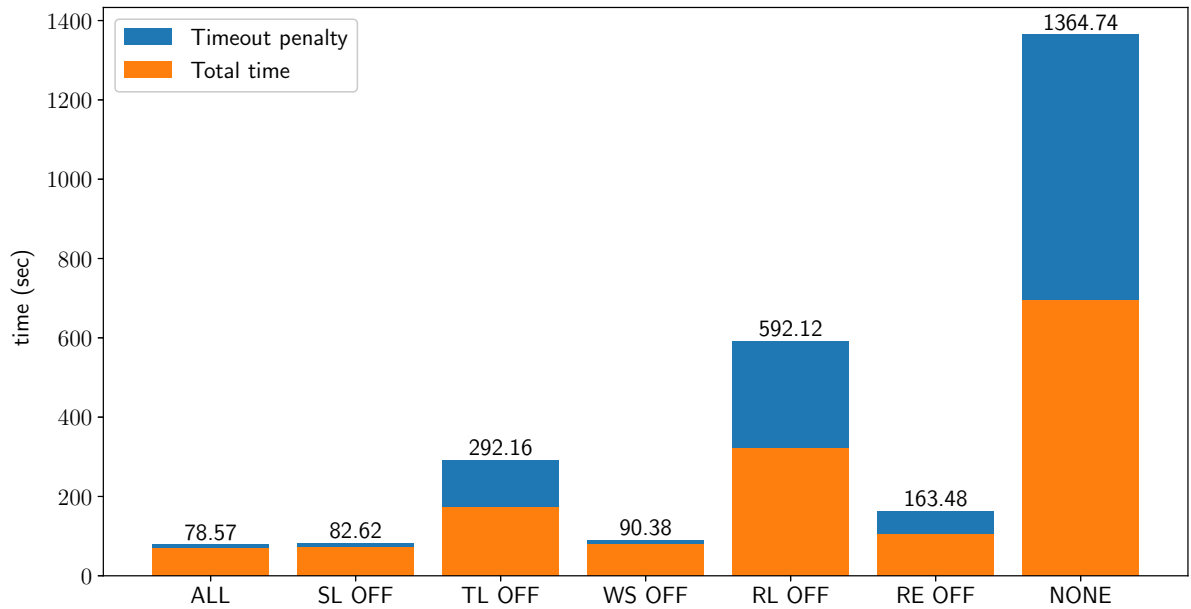}
  \caption{The total time of all pruning tests}
  \label{fig:prune-times-comp}
\end{figure}

The summary of results is displayed in Figure \ref{fig:prune-times-comp} which shows the amount of time for each of the seven cases across 80 tests. The graph shows the sum of the measured times and also highlights penalties for timeouts. The smaller the individual bars are, the better the result. But in the case of the bars representing a specific pruning heuristic being turned off, the bigger the bar, the more important the given heuristic is because the result is that much worse without it.

From these results, it is clear that the tree pruning is the key feature of the tree search algorithm. After turning off the pruning, the results are quite poor---67 out of 80 tests timed out. The total time is then not relevant at all. The middle bars, which represent the individual pruning heuristics being turned off, need to be compared to the first one, where all heuristics are active, to see how important the given heuristic actually is. Thus, Figure \ref{fig:prune-times-comp} suggests that the RL (complementarity relation) check is the most important one because turning it off had the biggest impact on the result. This can be a bit misleading, as some heuristics can sometimes be backed up by another one. This is the reason why the WS (match of leftmost terminals to the input string) seems to have a rather small impact. If this heuristic is turned off, the dead branch can be identified by the RE (regular expression) check and so the impact is not so big. Similarly, turning off the SL (strands length) heuristic has a smaller impact because it is backed up by TL (total length) heuristic.

Nevertheless, all heuristics are useful according to this result because no other result is as good as in the first case, where all the heuristics are active.  This finding is especially important in the case of the RE heuristic. This one is quite demanding with regards to computational power---regular expression match is performed each time this check is executed. It is the reason why it is the last check that is used, if there is another heuristic able to prune the node, a lot of computational power is saved. However, Figure~\ref{fig:prune-times-comp} shows clearly that RE heuristic contributes significantly to the overall performance. Still, some tests cannot benefit from all heuristics and turning some off would improve the results. This, again, means that there is some space for improving the performance by customizing the algorithm to specific grammars.

\subsection{The time complexity of the state space search}
The previous sections showed that the best overall performance is achieved when using all of the tree pruning heuristics and using the NTA+TM1 as the node precedence heuristic. This may not be the case for every grammar or every input, but it is the case overall. Therefore, this will be the setting used in the sections that follow---testing the tree search performance, analyzing the practical complexity and comparison to the WK-CYK algorithm.

In order to test the complexity of the tree search, 80 test have been run, two for each of the 40 grammars---one with inputs that are going to be accepted by the tree search and one with inputs which will be rejected. Each test runs the tree search several times and increases the input length. It stops when the computation takes longer than a limit of ten seconds or after 30 runs.

\begin{figure}[h!]
  \includegraphics[width=\textwidth]{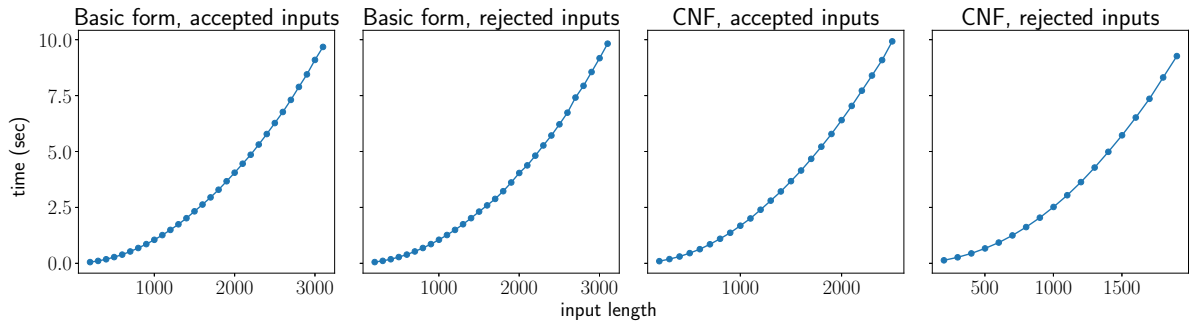}
  \caption{Grammar 12: $r^n d^n u^n r^n$}
  \label{fig:speed-test-grammar-12}
\end{figure}

The performance for different grammars is quite different. In the case of 11 of the 20 grammars (specifically, grammars 2, 5--10, 12-14, 19), the resulting graph is a very regular parabola. Often a bit steeper when transformed to the WK-CNF and often steeper when the inputs are going to be rejected. An example is in Figure \ref{fig:speed-test-grammar-12}. The performance with these grammars allows at least hundreds, in most cases thousands, of symbols on the input.

\begin{figure}[h!]
  \includegraphics[width=\textwidth]{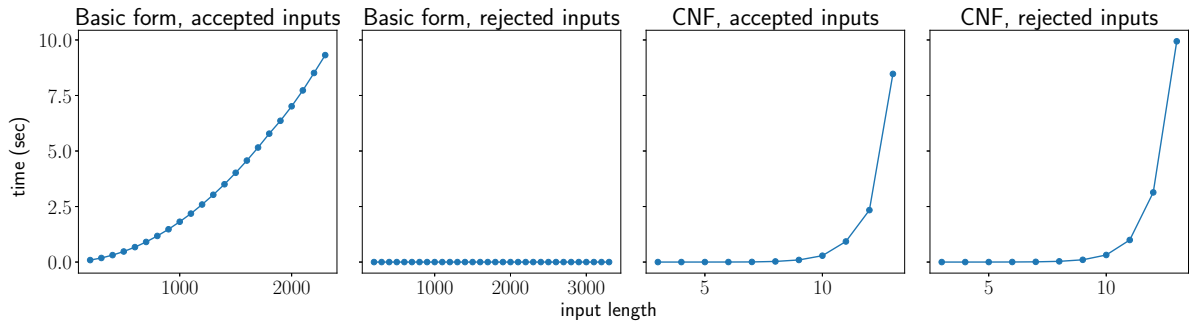}
  \caption{Grammar 3: $(a+b+c)^*abc$ with left recursive rules}
  \label{fig:speed-test-grammar-3}
\end{figure}

\begin{figure}[h!]
  \includegraphics[width=\textwidth]{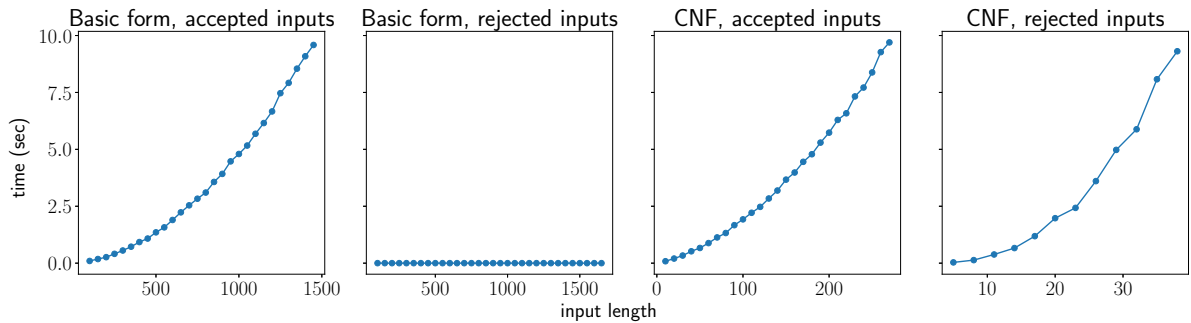}
  \caption{Grammar 4: $(a+b+c)^*abc$ with right recursive rules}
  \label{fig:speed-test-grammar-4}
\end{figure}

Occasionally, thanks to the pruning heuristics, the algorithm is able to tell practically immediately that there is no solution. This is the case with grammar 3 (Figure \ref{fig:speed-test-grammar-3}) and grammar 4 (Figure \ref{fig:speed-test-grammar-4}) in basic forms, making the complexity of this particular search constant. The grammar 3 has as the first rule, which it has to use to proceed further, $S \rightarrow A \wkpair{abc}{abc}$. If the input string does not end with $abc$, the regular expression check immediately detects that the input cannot be matched, it prunes the only branch and the search is finished.

Similarly, in the case of grammar 4, any inputs that are longer than seven symbols need to end with the symbol $a$ and can be reached only by using $S \rightarrow Q \wkpair{a}{a}$ as the first rule. The regular expression check immediately prunes this branch. The rest of the tree is searched very quickly because the only other possible starting rule is $S \rightarrow A B C D E F G$, there are not many states that can be reached from it, so this part of the tree is always small.

After the conversion of the grammar to the WK-CNF, the complexity usually goes up. The transformation adds a lot more rules and so the state space expands more rapidly, there are also longer paths from the starting non-terminal to the final string (containing only terminals), making the tree deeper. Also, the node precedence heuristics and the pruning have a harder time because many rules contain non-terminals only and most of the heuristics work with terminals. The most extreme case is the grammar 3 (Figure \ref{fig:speed-test-grammar-4}) where the tree search is very effective for grammar in basic form (as discussed, in the case of rejecting inputs, the result is immediate) but has very bad effectiveness for this grammar in the WK-CNF. The maximum length of input it can answer within 10 seconds is about 13--14 symbols.

\begin{figure}[h!]
  \includegraphics[width=\textwidth]{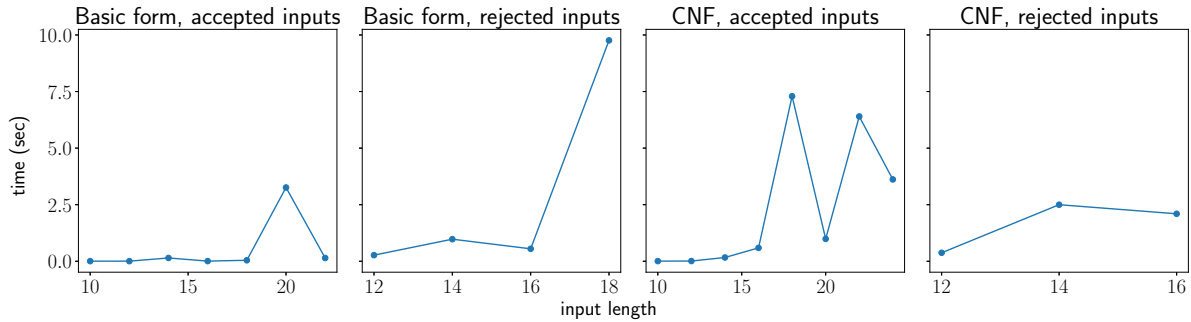}
  \caption{Grammar 17: $w: |w|_a = |w|_b$ and for any prefix $v$ of $w$: $v: |v|_a \geq |v|_b$ where $|x|_a$ denotes the number of occurrences of symbol $a$ in the string $x$}
  \label{fig:speed-test-grammar-17}
\end{figure}

The worst results for grammar in the basic form are in the case of grammar 17 (Figure \ref{fig:speed-test-grammar-17}). Here, the tree search can handle only inputs with the length of about 20 symbols within 10 seconds.

\begin{figure}[h!]
  \includegraphics[width=\textwidth]{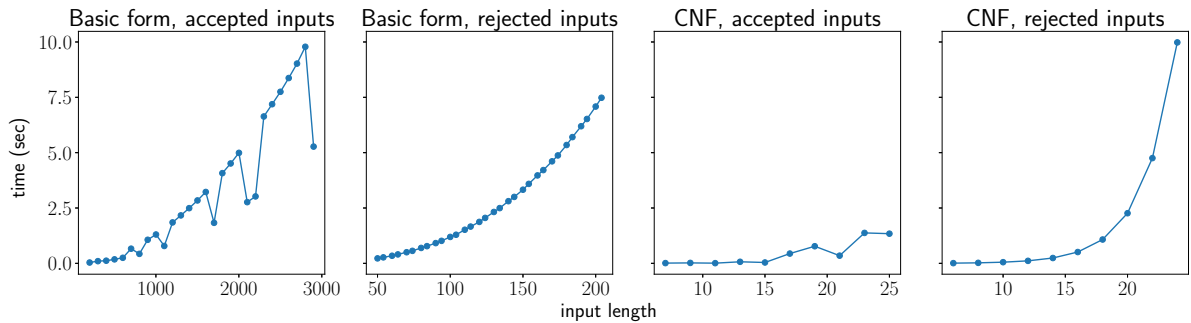}
  \caption{Grammar 11: $(ww)^C$}
  \label{fig:speed-test-grammar-11}
\end{figure}

\begin{figure}[h!]
  \includegraphics[width=\textwidth]{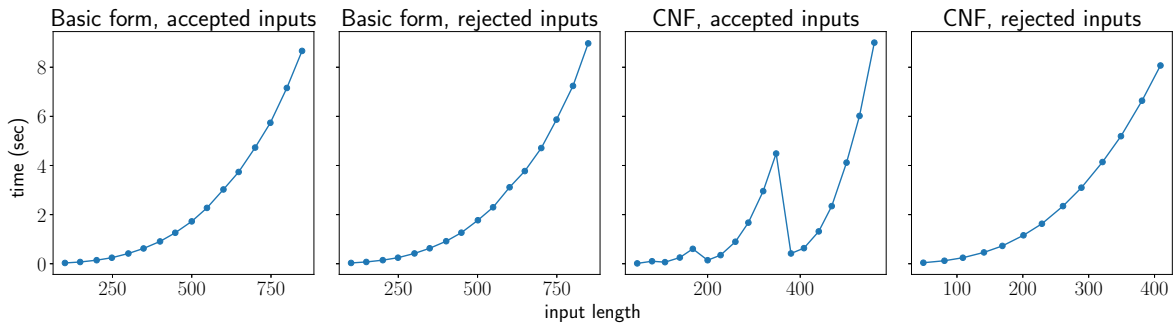}
  \caption{Grammar 16: $a^n b^m a^n$ where $2n \leq m \leq 3n$}
  \label{fig:speed-test-grammar-16}
\end{figure}

\begin{figure}[h!]
  \includegraphics[width=\textwidth]{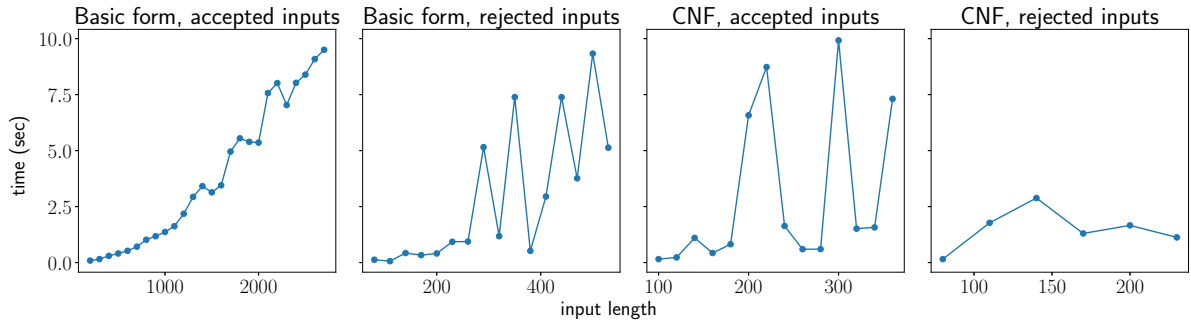}
  \caption{Grammar 18: $(l^n r^n)^k$ where $n$ does not increase for subsequent $k$s}
  \label{fig:speed-test-grammar-18}
\end{figure}

Some grammars manifest an interesting behavior---for some longer inputs, the performance is actually better. This is the case of grammars 11 (Figure \ref{fig:speed-test-grammar-11} on the left), 16 (Figure \ref{fig:speed-test-grammar-16} on the middle right) and 18 (Figure \ref{fig:speed-test-grammar-18} on the middle left and middle right). Some of these results may appear to be partly random, simply the input generator might sometimes generate an input which is more complex and sometimes less complex to compute. This is the case of the grammar 18, here, in fact, the shape of the randomly generated input has a significant impact on the performance which is the reason for the irregular curves. However, repeated tests of the other two grammars confirmed that this always happens and the figure of grammar 16 has a clear pattern, it is certainly not random.

The results of the three grammars which have not been yet mentioned, grammars 1, 15 and 20, are mostly similar to the standard parabola of the majority of tests with some irregularities.

\subsection{Testing the efficiency of WK-CYK}
The WK-CYK algorithm has been tested in a similar manner as the tree search. This time not all grammars can be used due to the limitations of WK-CYK. The grammars must be in the WK-CNF and grammars 5, 19 and 20 cannot be used at all, since WK-CYK requires the complementarity relation to be identity which is not the case for these three grammars. Therefore there are 17 grammars that can be used---each is used for two tests, one with inputs that should be accepted and one with inputs that should be rejected. Again, each test increases the input string length until the computation lasts more than the limit of 10 seconds.

\begin{figure}[h!]
  \centering
  \includegraphics[width=0.70\textwidth]{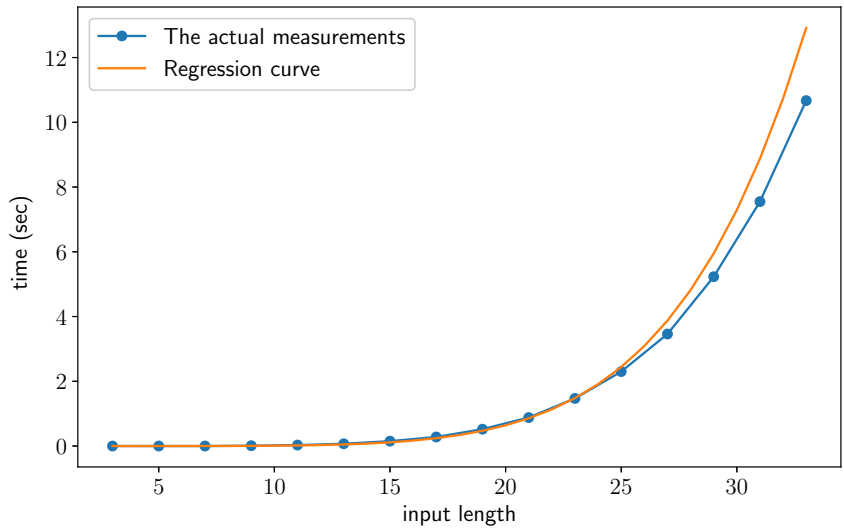}
  \caption{WK-CYK test results compared to the regression curve $x^6 / 10^8$}
  \label{fig:wk-cyk-test}
\end{figure}

It turns out that the WK-CYK gives very similar performance in all tests---for all the grammars and regardless of whether the input is accepted or not. Figure \ref{fig:wk-cyk-test} on the left shows a result for the first test which is very similar to all the others. The limit of ten seconds is reached by WK-CYK when the input has about 33 symbols. The graph also includes a regression curve $x^6 / 10^8$. These results agree with the claim made by the authors of WK-CYK that the complexity with regards to the input length is $\mathcal{O}(n^6)$.

When the results of the WK-CYK and state space search are compared, the advantage of state space search is the actual speed in most cases. The results in the previous sections showed that of all the grammars only one (grammar 17) was slower in the basic form when analyzed by tree search than when analyzed by WK-CYK. After transformation to WK-CNF two more grammars (grammar 3 and 11) were comparable or slower when analyzed by the tree search. Grammar 1 was slower for the negative inputs.

\subsection{Conclusion and possibilities for improvements}

It has been concluded in the previous section that the WK-CYK algorithm is able to compute within the time limit of ten seconds results for inputs of length of approximately 33 symbols. I assume that this will always be the case, even for grammars that would have to be modified in order to be suitable for WK-CYK (grammars 5, 19 and 20). If these results are compared with the results of state space search over grammars in basic forms, only one of them is more efficient with WK-CYK. Other 19 are more efficient with the state space search allowing hundreds of input symbols at minimum. That means that state space search was more efficient in 38 out of 40 test cases (each grammar is tested with accepted and rejected inputs), i.e., in 97.5 \% of cases. In this comparison, the state space search benefits from being able to work with any WK grammar---there is no need to transform it to the WK-CNF.

If the state space search is compared to WK-CYK over all 40 grammars, WK-CYK has better efficiency in case of grammars 3 and 11 (in the WK-CNF) and with the rejected inputs for grammar 1. That is 9 test cases out of 80 (four test cases of grammar 17, two of grammars 3 and 11, one of grammar 1), i.e., in 88,75 \% of cases. However, this comparison assumes that there is a need to use the grammars in the CNF.

These results show some advantages and disadvantages of the two algorithms. An advantage of state space search is the flexibility regarding the grammars. It does not require to work with grammars in the WK-CNF. Also, it does not require the complementarity relation to be identity. Even though it is always possible to transform any WK grammar to the WK-CNF and it is always possible to further transform the grammar in order to use only the identity as the relation, this can significantly add to the grammar's complexity.

A useful feature of the state space search is the fact that it can be configured for the need of a specific grammar. If the membership test will be performed repeatedly on a grammar, it is possible to find out what node precedence heuristic works best and what pruning heuristics are useful in that particular scenario by running tests analogous to those presented in Section \ref{section:node-prec-test}. Thus the performance may be further enhanced.

An advantage of WK-CYK, on the other hand, is its universality. It has roughly the same speed every time, it does not significantly depend on the grammar (increasing the number of rules adds a little bit) and it does not matter, if the input is going to be accepted or not. For very complicated grammars, especially with lots of rules or long derivations from the starting symbol to the final string, WK-CYK still might be more practical.

The state space search is a suitable algorithm for parallelization. Several processes can take nodes from the queue of open nodes and analyze different branches of the tree independently. This would be a natural next step in the further development of the state space search.
It is possible to come up with other heuristics for both the pruning and node precedence. As for pruning, one possibility would be to expand the regular expression matching (RE) heuristic to consider also the lower strand.
Another idea for pruning is to calculate how many terminals can be generated at minimum to the lower strand and to the upper strand individually (currently, it is calculated how many terminals a non-terminal produces to both strands) thus making the constraint of the WK-words stronger.
As for the node precedence heuristics, it may be worthwhile to use some of the grammars with which state space search is not efficient (in particular grammars 3 in the WK-CNF and grammar 17) and design or improve node precedence heuristics with respect to these particular cases. Then it would be necessary to test all these new heuristics and see if they contribute to the overall performance or not.
Another promising improvement could be analyzing the input from both sides at the same time. This could help with the cases, when the key part of the input is at or near its end and the state space search may struggle to get there in a reasonable time frame.

\section*{Acknowledgment}
\small
This work was supported by the~Ministry of Education, Youth and~Sports of Czech Republic project ERC.CZ no. LL1908 and the~BUT grant FIT-S-20-6293.

\nocite{*}
\bibliographystyle{eptcs}
\bibliography{generic}

\begin{thebibliography}{10}
\providecommand{\bibitemdeclare}[2]{}
\providecommand{\surnamestart}{}
\providecommand{\surnameend}{}
\providecommand{\urlprefix}{Available at }
\providecommand{\url}[1]{\texttt{#1}}
\providecommand{\href}[2]{\texttt{#2}}
\providecommand{\urlalt}[2]{\href{#1}{#2}}
\providecommand{\doi}[1]{doi:\urlalt{https://doi.org/#1}{#1}}
\providecommand{\eprint}[1]{arXiv:\urlalt{https://arxiv.org/abs/#1}{#1}}
\providecommand{\bibinfo}[2]{#2}

\bibitemdeclare{article}{WK-PUSHDOWN}
\bibitem{WK-PUSHDOWN}
\bibinfo{author}{Kingshuk \surnamestart Chatterjee\surnameend} \&
  \bibinfo{author}{Kumar~Sankar \surnamestart Ray\surnameend}
  (\bibinfo{year}{2017}): \emph{\bibinfo{title}{{Watson-Crick} Pushdown
  Automata}}.
\newblock {\slshape \bibinfo{journal}{Kybernetika}}
  \bibinfo{volume}{53}(\bibinfo{number}{5}), pp. \bibinfo{pages}{868--876},
  \doi{10.14736/kyb-2017-5-0868}.

\bibitemdeclare{article}{SURVEY}
\bibitem{SURVEY}
\bibinfo{author}{Elena \surnamestart Czeizler\surnameend} \&
  \bibinfo{author}{Eugen \surnamestart Czeizler\surnameend}
  (\bibinfo{year}{2006}): \emph{\bibinfo{title}{A Short Survey on
  {Watson-Crick} Automata}}.
\newblock {\slshape \bibinfo{journal}{Bull. {EATCS}}} \bibinfo{volume}{88}, pp.
  \bibinfo{pages}{104--119}.

\bibitemdeclare{inproceedings}{WK-FIN-AUT}
\bibitem{WK-FIN-AUT}
\bibinfo{author}{Rudolf \surnamestart Freund\surnameend},
  \bibinfo{author}{Gheorghe \surnamestart Paun\surnameend},
  \bibinfo{author}{Grzegorz \surnamestart Rozenberg\surnameend} \&
  \bibinfo{author}{Arto \surnamestart Salomaa\surnameend}
  (\bibinfo{year}{1997}): \emph{\bibinfo{title}{{Watson-Crick} Finite
  Automata}}.
\newblock In \bibinfo{editor}{Harvey \surnamestart Rubin\surnameend} \&
  \bibinfo{editor}{David~Harlan \surnamestart Wood\surnameend}, editors:
  {\slshape \bibinfo{booktitle}{DNA Based Computers, Proceedings of a DIMACS
  Workshop, Philadelphia, Pennsylvania, USA, June 23-25, 1997}}, {\slshape
  \bibinfo{series}{DIMACS Series in Discrete Mathematics and Theoretical
  Computer Science}}~\bibinfo{volume}{48}, \bibinfo{publisher}{DIMACS/AMS}, pp.
  \bibinfo{pages}{297--328}, \doi{10.1090/dimacs/048/22}.

\bibitemdeclare{mastersthesis}{DIPL}
\bibitem{DIPL}
\bibinfo{author}{Jan \surnamestart Hammer\surnameend} (\bibinfo{year}{2022}):
  \emph{\bibinfo{title}{{Watson-Crick} Models for Formal Language Processing}}.
\newblock \bibinfo{type}{Master's thesis}, \bibinfo{school}{Brno University of
  Technology, Faculty of Information Technology}.

\bibitemdeclare{article}{CYK}
\bibitem{CYK}
\bibinfo{author}{John~E. \surnamestart Hopcroft\surnameend},
  \bibinfo{author}{Rajeev \surnamestart Motwani\surnameend} \&
  \bibinfo{author}{Jeffrey~D. \surnamestart Ullman\surnameend}
  (\bibinfo{year}{2001}): \emph{\bibinfo{title}{Introduction to automata
  theory, languages, and computation, 2nd edition}}.
\newblock {\slshape \bibinfo{journal}{SIGACT news}}
  \bibinfo{volume}{32}(\bibinfo{number}{1}), pp. \bibinfo{pages}{60--65},
  \doi{10.1145/568438.568455}.

\bibitemdeclare{incollection}{COMPL-REL}
\bibitem{COMPL-REL}
\bibinfo{author}{Dietrich \surnamestart Kuske\surnameend} \&
  \bibinfo{author}{Peter \surnamestart Weigel\surnameend}
  (\bibinfo{year}{2004}): \emph{\bibinfo{title}{The Role of the Complementarity
  Relation in {Watson-Crick} Automata and Sticker Systems}}.
\newblock In: {\slshape \bibinfo{booktitle}{Developments in Language Theory}},
  \bibinfo{series}{Lecture Notes in Computer Science},
  \bibinfo{publisher}{Springer Berlin Heidelberg}, \bibinfo{address}{Berlin,
  Heidelberg}, pp. \bibinfo{pages}{272--283},
  \doi{10.1007/978-3-540-30550-7\_23}.

\bibitemdeclare{article}{WK-GRAMMARS-1}
\bibitem{WK-GRAMMARS-1}
\bibinfo{author}{Nurul~Liyana \surnamestart Mohamad~Zulkufli\surnameend},
  \bibinfo{author}{Sherzod \surnamestart Turaev\surnameend},
  \bibinfo{author}{Mohd~Izzuddin \surnamestart Mohd~Tamrin\surnameend} \&
  \bibinfo{author}{Azeddine \surnamestart Messikh\surnameend}
  (\bibinfo{year}{2016}): \emph{\bibinfo{title}{Generative Power and Closure
  Properties of {Watson-Crick} Grammars}}.
\newblock {\slshape \bibinfo{journal}{Applied computational intelligence and
  soft computing}} \bibinfo{volume}{2016}, pp. \bibinfo{pages}{1--12},
  \doi{10.1155/2016/9481971}.

\bibitemdeclare{article}{WK-CYK}
\bibitem{WK-CYK}
\bibinfo{author}{Nurul~Liyana \surnamestart Mohamad~Zulkufli\surnameend},
  \bibinfo{author}{Sherzod \surnamestart Turaev\surnameend},
  \bibinfo{author}{Mohd~Izzuddin \surnamestart Mohd~Tamrin\surnameend} \&
  \bibinfo{author}{Azeddine \surnamestart Messikh\surnameend}
  (\bibinfo{year}{2018}): \emph{\bibinfo{title}{{Watson-Crick} Context-Free
  Grammars: Grammar Simplifications and a Parsing Algorithm}}.
\newblock {\slshape \bibinfo{journal}{The Computer Journal}}
  \bibinfo{volume}{61}(\bibinfo{number}{9}), pp. \bibinfo{pages}{1361--1373},
  \doi{10.1093/comjnl/bxx128}.

\bibitemdeclare{inproceedings}{WK-GRAMMARS-2}
\bibitem{WK-GRAMMARS-2}
\bibinfo{author}{Nurul~Liyana \surnamestart Mohamad~Zulkufli\surnameend},
  \bibinfo{author}{Sherzod \surnamestart Turaev\surnameend},
  \bibinfo{author}{Mohd Izzuddin~Mohd \surnamestart Tamrin\surnameend} \&
  \bibinfo{author}{Azeddine \surnamestart Messikh\surnameend}
  (\bibinfo{year}{2017}): \emph{\bibinfo{title}{The Computational Power of
  {Watson-Crick} Grammars: Revisited}}.
\newblock In: {\slshape \bibinfo{booktitle}{Bio-inspired Computing – Theories
  and Applications}}, {\slshape \bibinfo{series}{Communications in Computer and
  Information Science}} \bibinfo{volume}{681}, \bibinfo{publisher}{Springer
  Singapore}, \bibinfo{address}{Singapore}, pp. \bibinfo{pages}{215--225},
  \doi{10.1007/978-981-10-3611-8\_20}.

\bibitemdeclare{book}{DNA-computing}
\bibitem{DNA-computing}
\bibinfo{author}{Gheorghe \surnamestart Păun\surnameend},
  \bibinfo{author}{Grzegorz \surnamestart Rozenberg\surnameend} \&
  \bibinfo{author}{Arto \surnamestart Salomaa\surnameend}
  (\bibinfo{year}{1998}): \emph{\bibinfo{title}{DNA Computing: New Computing
  Paradigms}}.
\newblock \bibinfo{series}{Texts in Theoretical Computer Science. An EATCS
  Series}, \bibinfo{publisher}{Springer Berlin Heidelberg},
  \bibinfo{address}{Berlin, Heidelberg}, \doi{10.1007/978-3-662-03563-4}.

\bibitemdeclare{inproceedings}{REG-GRAMMAR}
\bibitem{REG-GRAMMAR}
\bibinfo{author}{K.~G. \surnamestart Subramanian\surnameend},
  \bibinfo{author}{S.~\surnamestart Hemalatha\surnameend} \&
  \bibinfo{author}{Ibrahim \surnamestart Venkat\surnameend}
  (\bibinfo{year}{2012}): \emph{\bibinfo{title}{On {Watson-Crick} Automata}}.
\newblock In: {\slshape \bibinfo{booktitle}{Proceedings of the Second
  International Conference on Computational Science, Engineering and
  Information Technology}}, \bibinfo{series}{CCSEIT '12},
  \bibinfo{publisher}{Association for Computing Machinery},
  \bibinfo{address}{New York, NY, USA}, p. \bibinfo{pages}{151–156},
  \doi{10.1145/2393216.2393242}.

\end{thebibliography}

\appendix
\newpage

\section{The WK-CYK algorithm} \label{app:wk-cyk}
\begin{lstlisting}[caption={Procedure SetsConstruction of WK-CYK}, label={code:wk-cyk-main}, escapeinside={(*}{*)},numbers=left,
numberstyle=\small, numbersep=8pt,frame = single, framexleftmargin=15pt]
procedure SetsConstruction:
Input: string [(*$w/w$*)] = [(*$x_{11}x_{12}...x_{1n} / x_{21}x_{22}...x_{2n}$*)]

for (*1 $\leq i \leq n$*) do
    (*$X_{i:i,0:0} = \{A: A \rightarrow (x_{1i}/\lambda)\}$*)
    (*$X_{0:0,i:i} = \{A: A \rightarrow (\lambda/x_{2i})\}$*)

for (*$2 \leq y \leq 2n $*) do
    for (*$0 \leq \beta \leq n$*) do
        (*$\alpha = y - \beta$*)
        if (*$\alpha = 0$*) then
            i = j = 0
            for (*$1 \leq k \leq n - y + 1$*) do
                l = k + y - 1
                ComputeSet (*$X_{i:j,k:l}$*)
        else if (*$\beta = 0$*) then
            k = l = 0
            for (*$1 \leq i \leq n - y + 1$*) do
                j = i + y - 1
                ComputeSet (*$X_{i:j,k:l}$*)
        else
            for (*$1 \leq i \leq n - \alpha + 1$*) do
                for (*$1 \leq k \leq n - \beta + 1$*) do
                    j = i + (*$\alpha$*) - 1
                    l = k + (*$\beta$*) - 1
                    ComputeSet (*$X_{i:j,k:l}$*)

(*$w \in L(G)$*) iff (*$S \in X_{1:n,1:n}$*)
\end{lstlisting}

\begin{lstlisting}[caption={Procedure ComputeSet of WK-CYK}, label={code:wk-cyk-compute-sets}, escapeinside={(*}{*)},
numbers=left,numberstyle=\small, numbersep=8pt,frame = single, framexleftmargin=15pt]
procedure ComputeSet:

if (*$i = j$*) = 0 then
    (*$ X_{0:0,k:l} = \big\{\bigcup_{t \in [k, l-1]} X_{0:0,k:t} X_{0:0,t+1:l}\big\}$*)
else if (*$k = l$*) = 0 then
    (*$ X_{i:j,0:0} = \big\{\bigcup_{s \in [i, j-1]} X_{i:s,0:0} X_{s+1:j,0:0}\big\}$*)
else
    (*$ X_{i:j,k:l} = \big\{X_{i:j,0:0}X_{0:0,k:l} \cup X_{0:0,k:l}X_{i:j,0:0}\big\} \cup$*)
        (*$\bigcup_{s \in [i, j-1], t \in [k, l-1]} \big\{X_{i:s,k:t}X_{s+1:j,t+1:l}\big\} \cup$*)
        (*$\bigcup_{s \in [i, j-1]} \big\{X_{i:s,k:l}X_{s+1:j,0:0} \cup X_{i:s,0:0}X_{s+1:j,k:l}\big\} \cup$*)
        (*$\bigcup_{t \in [k, l-1]} \big\{X_{i:j,k:t}X_{0:0,t+1:l} \cup X_{0:0,k:t}X_{i:j,t+1:l}\big\}$*)
\end{lstlisting}

WK-CYK algorithm expects grammar in the WK-CNF and a double-stranded string on the input. Since the algorithm requires that the complementarity relation is the identity, the upper and lower strands are always the same.

WK-CYK uses sets marked as $X_{a:b,c:d}$. These are sets of non-terminals that can generate a segment of the input double-stranded string specified by the indexes $a$, $b$, $c$, and $d$. $a$ and $b$ are indexes of terminals in the upper strand and specify an interval (the indexing starts with index 1 and the edge indexes are included). The lower strand interval is specified by indexes $c$ and $d$. If a pair of indexes is 0, no symbols from the corresponding strand are included. 

For instance, for the segment $\wkpair{abcd}{abcd}$, $X_{2:2,0:0}$ would contain a set of non-terminals that generate $\wkpair{b}{\lambda}$, $X_{2:4,1:3}$ non-terminals that generate $\wkpair{bcd}{abc}$.

In the first step (lines 4--6 of Listing \ref{code:wk-cyk-main}) WK-CYK finds sets $X_{i:i,0:0}$ and $X_{0:0,i:i}$ for $1 \leq i \leq |n|$ ($n$ is the length of the input). These are non-terminals that directly generate single terminals. Then, it searches for ways to generate segments of the input of increasing lengths, beginning with length 2 and up to the length of $2n$. For the input of length $n$, there are $2n$ terminals---$n$ in the upper and $n$ in the lower strand. For each length of the segment, it takes all possible combinations of the number of symbols from the upper and the lower strands.

For each of these segments, the procedure \emph{ComputeSet} is called so it finds all non-terminals that can generate the given segment. When WK-CYK uses this procedure to compute set $X$ of a segment of length $n$, it is necessary to have already computed sets $X$ for all segments of length $m < n$. Therefore, it proceeds from length 1 upward.

When the segment of length $2n$ has been computed, WK-CYK is finished. It has succeeded if the starting symbol $S$ can generate the whole input; in other words: if $S \in X_{1:n,1:n}$.

The \emph{ComputeSet} procedure has as a parameter a segment of input specified by the four indexes. It searches all pairs of sets $X$ which could produce the given segment together. When all combinations of two $X$ sets potentially producing the input segment have been found, the procedure then needs to check the grammar rules to find those that generate non-terminal from these sets. This step is not explicitly described in the \emph{ComputeSet} procedure. For each such rule, the non-terminal on its left-hand side will be included in the procedure's result. The final result is then a set of all such non-terminals.

\section{Grammars used for testing}

The following Watson-Crick grammars have been used to test the tree search algorithm and the WK-CYK algorithm. Unless stated otherwise, the set of non-terminals and the set of terminals is defined simply by the symbols that appear in the rules---all the uppercase letters are non-terminals of the grammar, and all the lowercase letters and digits are terminals. The starting non-terminal is $S$, and the complementarity relation is identity. With these specifications in mind, the grammar can be defined just by the rules.

\begin{enumerate}
  \item{
    $$S \rightarrow \wkpair{a}{a} \:|\: S S S$$

    The accepted language is: $\{a\}\{aa\}^*$
  }

  \item{
    $$S \rightarrow \wkpair{a}{a} S \:|\: \wkpair{b}{b} S \:|\: \wkpair{c}{c} S \:|\: \wkpair{abc}{abc}$$

    The accepted language is: $\{a, b, c\}^*\{abc\}$

    This example aims to test inputs with the decisive part at the very end. This could prove difficult since the tree search expands the non-terminals from left to the right.
  }

  \item{
    $$S \rightarrow A \wkpair{abc}{abc}$$
    $$A \rightarrow A \wkpair{a}{a} \:|\: A \wkpair{b}{b} \:|\: A \wkpair{c}{c} \:|\: \wkpair{\lambda}{\lambda}$$

    The accepted language is: $\{a, b, c\}^*\{abc\}$

    The aim of this example is, again, to test inputs with the decisive part on the very end while, at the same time, the rules are left recursive.
  }

  \item{
    $$S \rightarrow Q \wkpair{a}{a} \:|\: A B C D E F G$$
    $$Q \rightarrow Q Q \:|\: A B C D E F G$$
    $$A \rightarrow \wkpair{a}{a} \:|\: \wkpair{\lambda}{\lambda}$$
    $$B \rightarrow \wkpair{b}{b} \:|\: \wkpair{\lambda}{\lambda}$$
    $$C \rightarrow \wkpair{c}{c} \:|\: \wkpair{\lambda}{\lambda}$$
    $$D \rightarrow \wkpair{d}{d} \:|\: \wkpair{\lambda}{\lambda}$$
    $$E \rightarrow \wkpair{e}{e} \:|\: \wkpair{\lambda}{\lambda}$$
    $$F \rightarrow \wkpair{f}{f} \:|\: \wkpair{\lambda}{\lambda}$$
    $$G \rightarrow \wkpair{g}{g} \:|\: \wkpair{\lambda}{\lambda}$$

    The accepted language is: ($a?b?c?d?e?f?g?) \cup (a?b?c?d?e?f?g?)^*\{a\}$ ($x?$ denotes that the symbol $x$ is optional, i.e. $(\{x\} \cup \lambda)$ )

    The problematic feature of this grammar may be the fact, that during the transformation of this grammar to the WK-CNF (more specifically, when removing the $\lambda$-rules) the number of rules increases rapidly.
  }

  \item{
    $$S \rightarrow \wkpair{a}{t} S \:|\: \wkpair{t}{a} S \:|\: \wkpair{g}{c} S \:|\: \wkpair{c}{g} A$$
    $$A \rightarrow \wkpair{c}{g} A \:|\: \wkpair{a}{t} S \:|\: \wkpair{g}{c} S \:|\: \wkpair{t}{a} B$$
    $$B \rightarrow \wkpair{c}{g} A \:|\: \wkpair{a}{t} S \:|\: \wkpair{t}{a} S \:|\: \wkpair{g}{c} C$$
    $$C \rightarrow \wkpair{a}{t} C \:|\: \wkpair{t}{a} C \:|\: \wkpair{g}{c} C \:|\: \wkpair{c}{g} C \:|\: \wkpair{\lambda}{\lambda}$$

     The terminals in this grammar refer to the actual nucleobases in the DNA and the complementarity relation mirrors the relations among them: $\rho = \{(a, t), (t, a), (c, g), (g, c)\}$

    The accepted language is: $(\{a,t,c,g\}^*\{ctg\}\{a,t,c,g\}^*)^*$

    This grammar is taken from \cite{WK-GRAMMARS-1} and is a first step towards an actual analysis of the DNA. In this case, it simply looks for the substring $ctg$.
  }

  \item{
    $$S \rightarrow \wkpair{a}{\lambda} S \:|\: \wkpair{a}{\lambda} A$$
    $$A \rightarrow \wkpair{b}{a} A \:|\: \wkpair{b}{a} B$$
    $$B \rightarrow \wkpair{\lambda}{b} B \:|\: \wkpair{\lambda}{b}$$

    The accepted language is: $a^nb^n$ where $n \geq 1$ (symbol $x^n$ denotes $n$ occurrences of the symbol $x$)

    The grammar is taken from \cite{REG-GRAMMAR}.
  }

  \item{
    $$S \rightarrow \wkpair{a}{a} S \wkpair{a}{a} \:|\: \wkpair{b}{b} S \wkpair{b}{b} \:|\: \wkpair{c}{c}$$

    The accepted language is: $wcw^R$ where $w \in \{a, b\}^*$ ($w^R$ is the reversal of the string $w$)
  }

  \item{
    $$S \rightarrow \wkpair{a}{a} S \wkpair{a}{a} \:|\: \wkpair{b}{b} S \wkpair{b}{b} \:|\: \wkpair{\lambda}{\lambda}$$

    The accepted language is: $ww^R$ where $w \in \{a, b\}^*$
  }

  \item{
    $$S \rightarrow B L \:|\: R B$$
    $$L \rightarrow B L \:|\: A$$
    $$R \rightarrow R B \:|\: A$$
    $$A \rightarrow B A B \:|\: \wkpair{2}{2}$$
    $$B \rightarrow \wkpair{0}{0} \:|\: \wkpair{1}{1}$$

    The accepted language is: $x2y: x, y \in \{0,1\}^* \wedge \:|x| \neq |y|$

    The grammar is taken from \footnote{Emanuele Viola (2016): \textit{Context-Free Languages}. \url{https://www.ccs.neu.edu/home/viola/classes/
slides/slides-context-free.pdf}. Accessed: 2022-10-05.}.
  }

  \item{
    $$S \rightarrow T \:|\: T \wkpair{p}{p} S$$
    $$T \rightarrow F \:|\: F T$$
    $$F \rightarrow \wkpair{e}{e} \:|\: W \:|\: \wkpair{o}{o} T \wkpair{p}{p} S \wkpair{c}{c} \:|\: X \wkpair{s}{s} \:|\: \wkpair{o}{o} Y \wkpair{c}{c} \wkpair{s}{s}$$
    $$X \rightarrow \wkpair{e}{e} \:|\: \wkpair{l}{l} \:|\: \wkpair{0}{0} \:|\: \wkpair{1}{1}$$
    $$Y \rightarrow T \wkpair{p}{p} S \:|\: F \wkpair{d}{d} T \:|\: X \wkpair{s}{s} \:|\: \wkpair{o}{o} Y \wkpair{c}{c} \wkpair{s}{s} \:|\: Z Z$$
    $$W \rightarrow \wkpair{l}{l} \:|\: Z$$
    $$Z \rightarrow \wkpair{0}{0} \:|\: \wkpair{1}{1} \:|\: Z Z$$

    The accepted language includes regular expressions over symbols 0 and 1 with parentheses ($o$ for opening and $c$ for closing parenthesis), operators $+$ (p), $*$ (s), $\cdot$ (d) and symbols $\emptyset$ (e), $\varepsilon$ (l)

    The grammar is taken from \footnote{Jeff Erickson (2018): \textit{Context-Free Languages and Grammars}. \url{https://jeffe.cs.illinois.edu/
teaching/algorithms/models/05-context-free.pdf}. Accessed: 2022-10-05.}.
  }

  \item{
    $$S \rightarrow A \:|\: B \:|\: A B \:|\: B A$$
    $$A \rightarrow \wkpair{a}{a} \:|\: \wkpair{a}{a} A \wkpair{a}{a} \:|\: \wkpair{a}{a} A \wkpair{b}{b} \:|\: \wkpair{b}{b} A \wkpair{b}{b} \:|\: \wkpair{b}{b} A \wkpair{a}{a}$$
    $$B \rightarrow \wkpair{b}{b} \:|\: \wkpair{a}{a} B \wkpair{a}{a} \:|\: \wkpair{a}{a} B \wkpair{b}{b} \:|\: \wkpair{b}{b} B \wkpair{b}{b} \:|\: \wkpair{b}{b} B \wkpair{a}{a}$$

    The accepted language is: $\{a, b\}^* \setminus \{ww \where w \in \{a, b\}^*\}$ -- i.e. the complement of the copy language.
  }

  \item{
    $$S \rightarrow \wkpair{r}{\lambda} S \:|\: \wkpair{r}{\lambda} A$$
    $$A \rightarrow \wkpair{d}{r} A \:|\: \wkpair{d}{r} B$$
    $$B \rightarrow \wkpair{u}{d} B \:|\: \wkpair{u}{d} C$$
    $$C \rightarrow \wkpair{r}{u} C \:|\: \wkpair{r}{u} D$$
    $$D \rightarrow \wkpair{\lambda}{r} D \:|\: \wkpair{\lambda}{r}$$

    The accepted language is: $r^nd^nu^nr^n$ where $n \geq 1$

    The grammar is taken from \cite{REG-GRAMMAR}.
  }

  \item{
    $$S \rightarrow \wkpair{a}{\lambda} S \wkpair{b}{\lambda} \:|\: \wkpair{a}{\lambda} A \wkpair{b}{\lambda}$$
    $$A \rightarrow \wkpair{c}{a} A \:|\: \wkpair{\lambda}{c} B \wkpair{\lambda}{b}$$
    $$B \rightarrow \wkpair{\lambda}{c} B \wkpair{\lambda}{b} \:|\: \wkpair{\lambda}{\lambda}$$

    The accepted language is: $a^nc^nb^n$ where $n \geq 1$

    The grammar is taken from \cite{WK-GRAMMARS-1}.

  }

  \item{
    $$S \rightarrow \wkpair{a}{\lambda} S \:|\: \wkpair{a}{\lambda} A$$
    $$A \rightarrow \wkpair{b}{\lambda} A \:|\: \wkpair{b}{\lambda} B$$
    $$B \rightarrow \wkpair{c}{a} B \:|\: \wkpair{c}{a} C$$
    $$C \rightarrow \wkpair{d}{b} C \:|\: \wkpair{d}{b} D$$
    $$D \rightarrow \wkpair{\lambda}{c} D \:|\: \wkpair{\lambda}{d} D \:|\: \wkpair{\lambda}{\lambda}$$

    The accepted language is: $a^nb^mc^nd^m$ where $n, m \geq 1$

    The grammar is taken from \cite{WK-GRAMMARS-1}.
  }

  \item{
    $$S \rightarrow \wkpair{a}{\lambda} S \:|\: \wkpair{b}{\lambda} S \:|\: \wkpair{c}{\lambda} A$$
    $$A \rightarrow \wkpair{a}{a} A \:|\: \wkpair{b}{b} A \:|\: \wkpair{\lambda}{c} B$$
    $$B \rightarrow \wkpair{\lambda}{a} B \:|\: \wkpair{\lambda}{b} B \:|\: \wkpair{\lambda}{\lambda}$$

    The accepted language is: $wcw$ where $w \in \{a, b\}^*$

    The grammar is taken from \cite{WK-GRAMMARS-1}.
  }

  \item{
    $$S \rightarrow \wkpair{a}{\lambda} S \wkpair{a}{a} \:|\: \wkpair{a}{\lambda} A \wkpair{a}{a} $$
    $$A \rightarrow \wkpair{bb}{a} A \:|\: \wkpair{bbb}{a} A \:|\: \wkpair{\lambda}{b} B$$
    $$B \rightarrow \wkpair{\lambda}{b} B \:|\: \wkpair{\lambda}{\lambda}$$

    The accepted language is: $a^nb^ma^n$ where $2n \leq m \leq 3n$

    The grammar is taken from \cite{WK-GRAMMARS-1}.
  }

  \item{
    $$S \rightarrow S S \:|\: \wkpair{a}{a} S \wkpair{b}{b} \:|\: \wkpair{a}{\lambda} S \:|\: \wkpair{a}{\lambda} A$$
    $$A \rightarrow \wkpair{b}{a} A \:|\: \wkpair{b}{a} B \:|\: \wkpair{b}{a}$$
    $$B \rightarrow \wkpair{\lambda}{b} B \:|\: \wkpair{\lambda}{b} \:|\: B B \:|\: \wkpair{a}{a} S \wkpair{b}{b} \:|\: \wkpair{a}{\lambda} S \:|\: \wkpair{a}{\lambda} A$$

    The accepted language is: $w: |w|_a = |w|_b$ and for any prefix $v$ of $w$: $v: |v|_a \geq |v|_b$ where $|x|_a$ denotes the number of occurrences of symbol $a$ in the string $x$

    The grammar is taken from \cite{WK-CYK}.
  }

  \item{
    $$S \rightarrow \wkpair{l}{\lambda} S \:|\: \wkpair{l}{\lambda} A$$
    $$A \rightarrow \wkpair{r}{l} A \:|\: \wkpair{r}{l} B$$
    $$B \rightarrow \wkpair{l}{r} B \:|\: \wkpair{\lambda}{r} B \:|\: \wkpair{\lambda}{\lambda} \:|\: A$$

    The accepted language is: $(l^n r^n)^k$ where $n$ does not increase for subsequent $k$. For instance: $lllrrrlrlr$ is within the language, $llrrlllrrr$ is not.

    The grammar is taken from \cite{WK-GRAMMARS-1} (where it is stated that the language of this grammar is $(l^nr^n)^k$ for $n, k \geq 1$ which is not correct). The original symbols for opening and closing parenthesis have been replaced by letters $l$ (left parenthesis) and $r$ (right parenthesis).
  }

  \item{
    The grammar is identical to the grammar 13 with a difference in the complementarity relation. The relations between symbols $a, b$ and symbols $a, c$ are added. This means that the relation is: $\rho = \{(a, a)$, $(b, b)$, $(c, c)$, $(a, b)$, $(b, a)$, $(a, c)$, $(c, a)\}$

    The accepted language is: $a^n b^m c^n$ where $n, m \geq 1$
  }

  \item{
	The grammar is identical to the grammar 14 with a difference in the complementarity relation. The relation between symbols $a, b$ is added making the relation $\rho = \{(a, a), (b, b), (c, c), (d, d), (a, b), (b, a)\}$

	The accepted language is: $a^m b^n c^o d^p$ where $m, n, o, p \geq 1 \wedge m+n = o+p$
  }

\end{enumerate}
\end{document}